\documentclass[]{spie}  

\usepackage{graphicx,amsmath,amssymb,caption,subcaption}

\graphicspath{{figures/}}
\title{Zernike amplitude pupil apodization for vortex coronagraphy with obscured apertures} 

\author{Garreth J. Ruane\supit{a}, Mark R. Dennis\supit{b}, and Grover A. Swartzlander\supit{a}
\skiplinehalf
\supit{a}Chester F. Carlson Center for Imaging Science, Rochester Institute of Technology, \\54 Lomb Memorial Drive, Rochester, NY 14623; \\
\supit{b}H. H. Wills Physics Laboratory, University of Bristol, Bristol BS8 1TL, United Kingdom
}
\authorinfo{Further author information: send correspondence to gjr8334@rit.edu}

\begin{document} 
\maketitle 
\begin{abstract}
A set of pupil apodization functions for use with a vortex coronagraph on telescopes with obscured apertures is presented. We show analytically that pupil amplitudes given by real-valued Zernike polynomials offer ideal on-axis starlight cancellation when applied to unobscured circular apertures. The charge of the vortex phase element must be a nonzero even integer, greater than the sum of the degree and the absolute value of its azimuthal order of the Zernike polynomial. Zero-valued lines and points of Zernike polynomials, or linear combinations thereof, can be matched to obstructions in the pupils of ground-based telescopes to improve the contrast achieved by a vortex coronagraph. This approach works well in the presence of a central obscuration and radial support structures. We analyze the contrast, off-axis throughput, and post-coronagraph point spread functions of an apodized vortex coronagraph designed for the European Extremely Large Telescope (E-ELT). This technique offers very good performance on apertures with large obscuring support structures similar to those on future 30-40m class ground-based telescopes. 
\end{abstract}


\section{INTRODUCTION}
The vortex coronagraph (VC) is a state-of-the-art instrument for extreme high-contrast astronomical imaging \cite{Mawet2005,Foo2005}. Recent on-sky observations with VC's on ground-based telescopes confirm that it is one of the best performing coronagraphic systems available, offering detection of dim companions at small angular separation from a star \cite{Mawet2010,Serabyn2010,Mawet2011a,Mawet2013a,Defrere2014}. Laboratory demonstrations have achieved contrast on the order of $10^{-9}$ at a few diffracted beamwidths, which may allow for direct detection and characterization of terrestrial exoplanets orbiting nearby stars \cite{Serabyn2013}.  

One practical problem is that many telescope pupils have secondary mirror and other obstructing features, such as spiders and discontinuities between mirror segments, that degrade the starlight suppression capability of coronagraphs. Several solutions have been developed for the VC to remedy the central obscuration owing to the secondary mirror including the use of a sub-aperture \cite{Mawet2010,Serabyn2010,Mawet2011a,Serabyn2007,Ruane2013}, tandem coronagraphic stages \cite{Mawet2011b, Mawet2013b}, interferometers \cite{Riaud2014}, and ring-shaped pupil apodizers \cite{Mawet2013b,Mawet2013c}. In addition, spiders and aperture discontinuities may be compensated for by numerically optimized binary amplitude pupil apodizers \cite{Carlotti2011,Carlotti2013,Carlotti2014}, continuous focal plane phase correcting elements \cite{Ruane2015}, or lossless apodization with a pair of deformable mirrors \cite{Pueyo2013,Pueyo2014,Fogarty2014}. 

We present a set of apodizers that offer a simple means to achieve high contrast in the presence of a central obscuration and radial spiders. The pupil functions are given by real-valued Zernike polynomials in field amplitude. The degree and azimuthal order of the Zernike polynomials are limited by the topological charge of the focal plane vortex phase element. This work represents a large expansion of the known analytical solutions for VC apodizers. The pupil functions described herein may be implemented by fabricating pupil-plane optical elements with complex transmittance, or possibly by use of lossless apodization techniques \cite{Guyon2003,Pueyo2013,Pueyo2014,Fogarty2014}.

\newpage
\section{VORTEX CORONAGRAPH WITHOUT APODIZATION}

\begin{figure}[t]
\begin{center}
\begin{tabular}{c}
\includegraphics{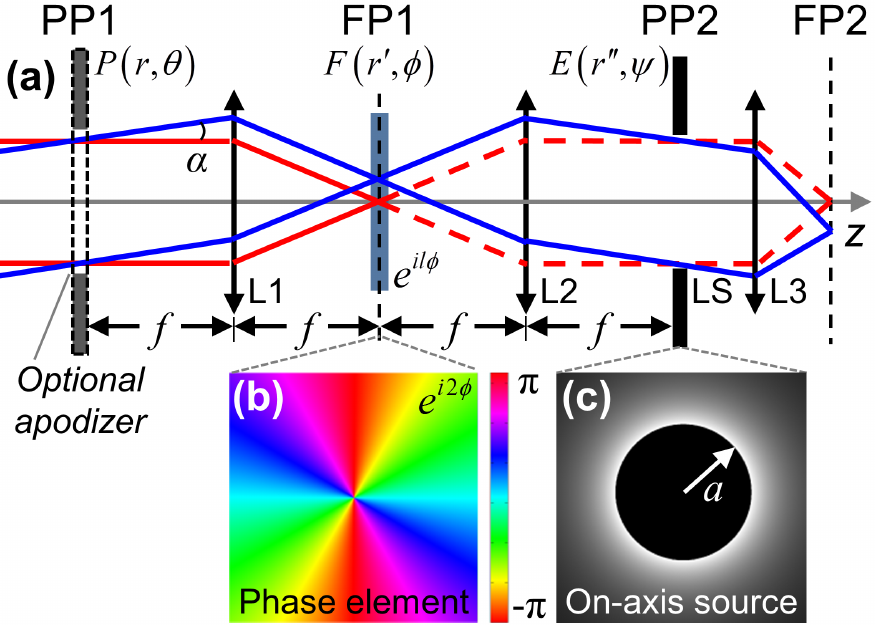} \hfill \includegraphics{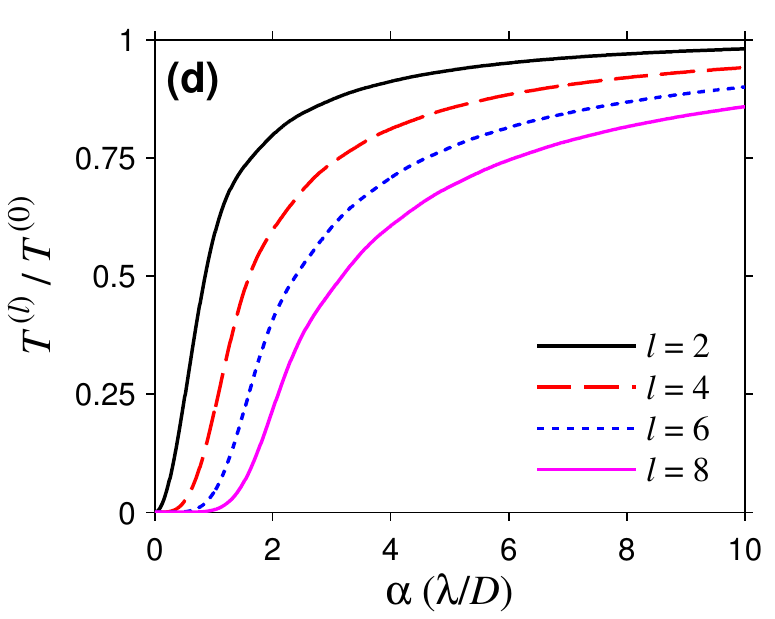}
\end{tabular}
\end{center}
\caption{ \label{fig:1} 
(a) Schematic of a vortex coronagraph. The first pupil plane (PP1) is typically formed by the reimaged telescope pupil. The field at PP1 is denoted $P(r,\theta)$, which may include an apodizer. Lens L1 forms the first focal plane (FP1), where a vortex phase element applies transmission function $\exp(il\phi)$. Lens L2 forms a second pupil plane (PP2), where an aperture known as the Lyot stop (LS) truncates the field. Lens L3 forms an image in focal plane (FP2) with the on-axis starlight removed. Light rays representing on-axis (red) and off-axis (blue) sources are drawn. (b) The phase pattern of an $l = 2$ vortex phase element. (c) Field magnitude just before the LS for a distant point source that coincides with the $z$ axis. (d) The relative LS throughput $T^{(l)}/T^{(0)}$ (with LS radius 0.97 times that of AP) for a point source at angular displacement $\alpha$ from the $z$ axis shown in units of $\lambda/D$, where $D=2a$ (i.e. the pupil diameter).}
\end{figure}

The basic operation of the VC is as follows: A vortex phase element located in the focal plane of a Lyot-style coronagraph diffracts on-axis starlight outside of a downstream aperture stop, while allowing light from off-axis sources to reach the subsequent image plane (see Fig. \ref{fig:1}) \cite{Mawet2005,Foo2005}. Mathematically, however, the formation of the circular ``nodal area" in the starlight at the exit pupil is not intuitive and is briefly derived here. 

As shown in Fig. \ref{fig:1}(a), the VC is a 4-$f$ optical system with a focal plane vortex phase element that has transmission $t(\phi)=\exp(il\phi),$ where $l$ is a nonzero even integer known as ``topological charge" and $\phi$ is the azimuthal angle in the first focal plane (FP1). Assuming an unobscured circular pupil with radius $a$ and without apodization, the field directly after the phase element owing to an on-axis point source may be written as
\begin{equation}
F\left(r',\phi \right)=\frac{k a^2}{f}\frac{J_1\left( k a r'/f\right)}{k a r'/f} e^{il\phi},
\label{eq1}
\end{equation}
where $r'$ is the radial coordinate in FP1, $k = 2\pi/\lambda,$ $\lambda$ is the wavelength, $f$ is the focal length, $J_1$ is the Bessel function of the first kind, and constant phase factors have been dropped. The exit pupil (PP2) field is given by the Fourier transform of Eq. \ref{eq1}:
\begin{equation}
E\left(r'',\psi \right)=e^{il\psi}\frac{ka}{f}W_1^l(r''),
\end{equation}
where $\left( r'', \psi \right)$ are the polar coordinates in PP2 and 
\begin{equation}
W_p^q(r'')=\int\limits_{0}^{\infty }{J_p\left({k a r'}/{f}\right)J_q\left( {k r'' r'}/{f}\right)dr'}.
\label{eq:Wpq}
\end{equation}
Eq. \ref{eq:Wpq} is related to the discontinuous integral of Weber and Schafheitlin \cite{Watson1922}. For $l$ is nonzero and even,
\begin{equation}
E\left(r'',\psi \right)=\left\{ \begin{matrix}
   0 & r'' \le a  \\
   \frac{a}{r''}R_{|l|-1}^1\left( \frac{a}{r''} \right){e^{il\psi}} & r''>a  \\
\end{matrix} \right. ,
\end{equation}
where $R_n^1$ are the radial Zernike polynomials (see Appendix \ref{sec:Zernikes} for definition) \cite{Carlotti2009}. The phase element and resulting field amplitude for $l=2$ are shown in Fig. \ref{fig:1}(b)-(c). The field is zero-valued within the geometric image of the pupil; that is, a nodal area appears. An aperture stop known as the ``Lyot stop" (LS), with radius $a_L$ where $a_L<a$, is placed in PP2 to block all of the light from the on-axis source. Off-axis sources do not form a nodal area and are partially transmitted through the LS. We define the throughput power for given value of $l$ as 
\begin{equation}
T^{(l)}=\int_0^{2\pi}\int_0^{a_L}\left|\tilde{E}\left(r'',\psi \right)\right|^2r''dr''d\psi,
\label{eq:throughputdef}
\end{equation}
where $\tilde{E}\left(r'',\psi \right)$ is the field at PP2 owing to a point source displaced from the optical axis by angle $\alpha$; i.e. at $\alpha=0$, $\tilde{E}\left(r'',\psi \right)=E\left(r'',\psi \right)$. The relative LS throughput with $a_L=0.97a$ is shown in Fig. \ref{fig:1}(d). For a non-apodized circular pupil $T^{(0)}/\mathcal{P}=\left(a_L/a\right)^2$, where $\mathcal{P}$ is the total power incident on PP2. 

\section{ZERNIKE AMPLITUDE APODIZED VORTEX CORONAGRAPH}
\begin{figure}[t!]
\begin{center}
\begin{tabular}{c}
\includegraphics{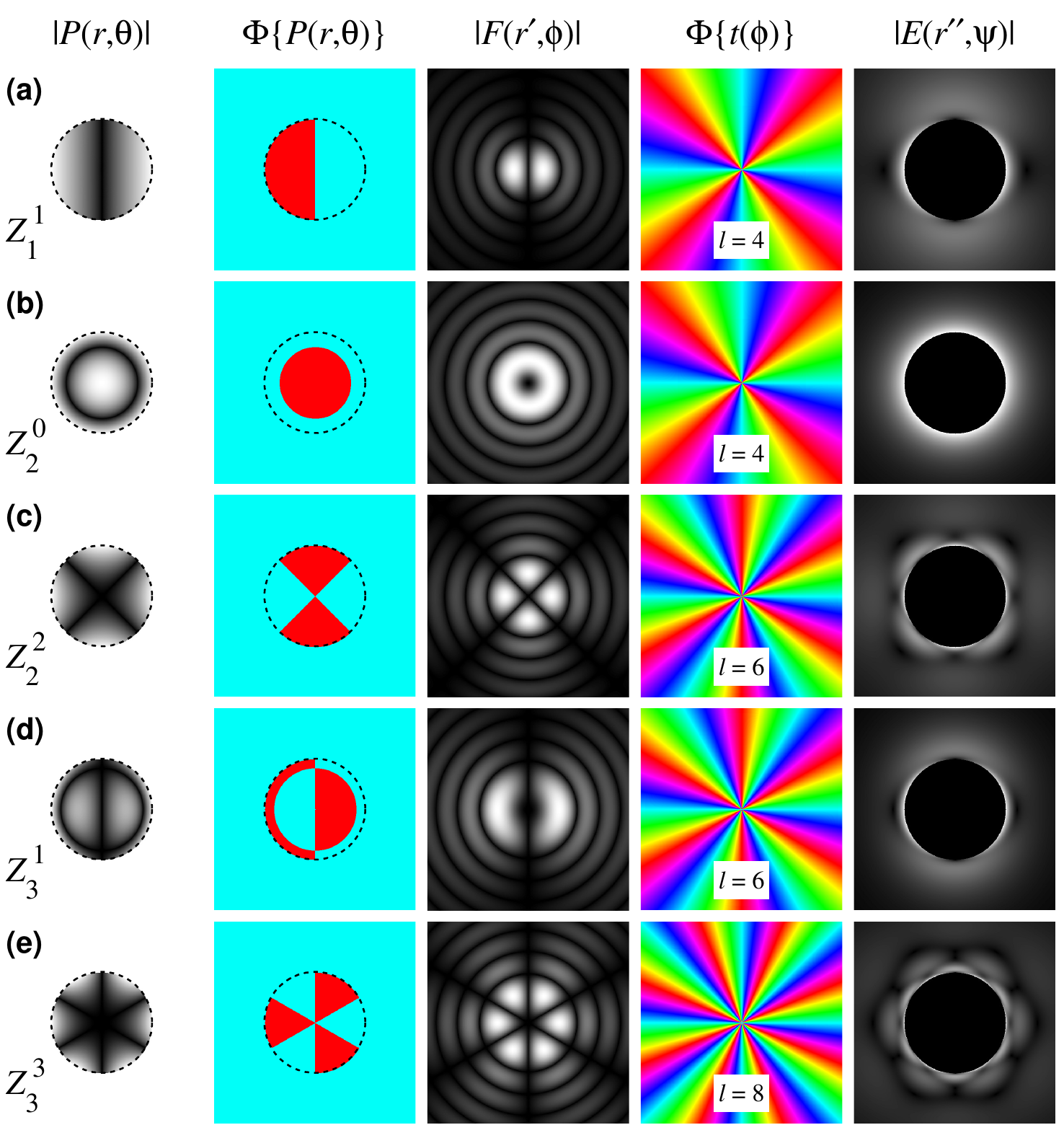} \hfill \\ 
    \begin{subfigure}{0.4\linewidth}
      \hspace{4mm}\includegraphics[trim = 0 0 0 33mm,clip=true]{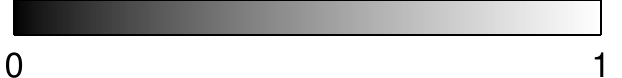}
    \end{subfigure}
    \hspace{10mm}
    \begin{subfigure}{0.4\linewidth}
      \includegraphics[trim = 0 0 0 31.5mm,clip=true]{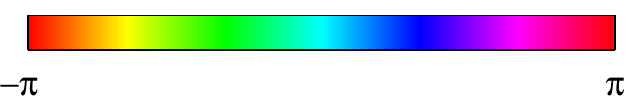}
    \end{subfigure}
\end{tabular}
\end{center}
\caption{ \label{fig:2} 
Zernike amplitude apodizers shown for (a) $Z_1^1$, (b) $Z_2^0$, (c) $Z_2^2$, (d) $Z_3^1$, and (e) $Z_3^3$ with an unobscured circular pupil of radius $a$ (dotted circle). For each case, the following are shown: the pupil amplitude $|P(r,\theta)|$ and phase $\Phi\{P(r,\theta)\}$ in PP1, the corresponding point spread function magnitude in FP1 $|F(r',\phi)|$, the phase of the lowest charge vortex phase element in FP1 that produces a nodal area in PP2 $\Phi\{t(\phi)\}$, and the field magnitude just before the LS in PP2 $|E\left(r'',\psi \right)|$. $|F(r',\phi)|$ is shown over a $10\times10 \; \lambda \, F\#$ square, where $F\#=f/(2a)$.}
\end{figure}

In this section, we introduce pupil amplitude functions described by 
\begin{equation}
P\left(r,\theta\right) = Z_n^m\left(r/a,\theta\right),\;\;\;\;\;r\le a,
\end{equation}
where $Z_n^m\left(r/a,\theta\right)$ are the real-valued Zernike polynomials and show that under certain conditions ideal contrast is achieved (see Fig. \ref{fig:2}). For $m\ge0$ (i.e. the even Zernike polynomials),
\begin{equation}
P\left(r,\theta\right) = R_n^m\left(r/a,\theta\right) \cos\left(m\theta\right),\;\;\;\;\;r\le a.
\label{eq:Zpupil}
\end{equation}
The field transmitted through a vortex phase element of charge $l$ in FP1, owing to an on-axis point source, is given by the product of $\exp\left(il\phi\right)$ and the Fourier transform of Eq. \ref{eq:Zpupil}:
\begin{equation}
F\left(r',\phi \right)=\frac{k a^2}{f}\frac{J_{n+1}\left( k a r'/f\right)}{k a r'/f} \cos\left(m\phi\right) e^{il\phi}.
\label{eq:PSF}
\end{equation}
The field in PP2 (just before the LS) is given by the Fourier transform of Eq. \ref{eq:PSF}:
\begin{equation}
E\left(r'',\psi \right)=\frac{k a}{2 f}e^{il\psi}\left[(-1)^m e^{im\psi}W_{n+1}^{l+m}(r'')+e^{-im\psi}W_{n+1}^{l-m}(r'')\right].
\label{eq:Wevens}
\end{equation}
Similarly, for $m<0$ (i.e. the odd Zernike polynomials)
\begin{equation}
E\left(r'',\psi \right)=\frac{k a}{i2 f}e^{il\psi}\left[(-1)^m e^{im\psi}W_{n+1}^{l+m}(r'')-e^{-im\psi}W_{n+1}^{l-m}(r'')\right].
\label{eq:Wodds}
\end{equation}
For an on-axis point source, a nodal area appears at PP2 if $|l|>n+|m|+1$ and $l$ is even valued. For $m=0$ and $|l|\le n$, the field at PP2 may be expressed
\begin{equation}
E\left(r'',\psi \right)=\left\{ \begin{matrix}
   R_n^l\left(r''/a\right)e^{il\psi} & r'' < a  \\
   0 & r''\ge a  \\
\end{matrix} \right. ,
\end{equation}
On the other hand, for $m=0$ and $|l|\ge n+2$
\begin{equation}
E\left(r'',\psi \right)=\left\{ \begin{matrix}
   0 & r'' \le a  \\
   \frac{a}{r''}R_{|l|-1}^{n+1}\left( \frac{a}{r''} \right){e^{il\psi}} & r''>a  \\
\end{matrix} \right. .
\end{equation}
More generally, analytical solutions for $E\left(r'',\psi \right)$ with $m\ne0$ that contain a nodal area in the on-axis starlight are written
\begin{equation}
E\left(r'',\psi \right)=\left\{ \begin{matrix}
   0 & r'' \le a  \\
   g^{(l)}_{n,m}\left(r'',\psi \right) & r''>a  \\
\end{matrix} \right. .
\end{equation}
Figure \ref{fig:2} shows the field patterns at PP2 for the first few Zernike amplitude apodizers with the lowest value of $l$ that yields a nodal area. Analytical expressions for $g^{(l)}_{n,m}\left(r'',\psi \right)$ are given in Appendix \ref{sec:EPfields}.

\section{ZERNIKE AMPLITUDE APODIZER FOR E-ELT}

\begin{figure}[t!]
\begin{center}
\begin{tabular}{c}
\includegraphics{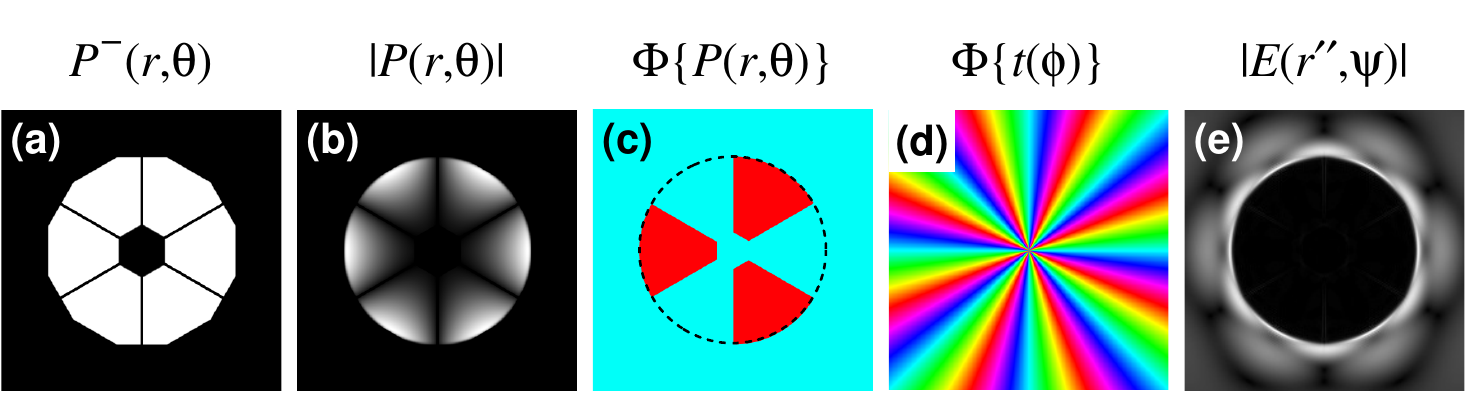}\hfill \\ 
    \begin{subfigure}{0.4\linewidth}
      \hspace{4mm}\includegraphics[trim = 0 0 0 33mm,clip=true]{colorbarsforfig1.pdf}
    \end{subfigure}
    \hspace{10mm}
    \begin{subfigure}{0.4\linewidth}
      \includegraphics[trim = 0 0 0 31.5mm,clip=true]{colorbarsforfig2.pdf}
    \end{subfigure}
\end{tabular}
\end{center}
\caption{ \label{fig:3} 
Pupil apodizer for a VC on the European Extremely Large Telescope (E-ELT). (a) The simplified pupil geometry matches well with the $Z_3^3$ Zernike polynomial (see Fig. \ref{fig:2}(e)). The pupil field just before the apodizer $P^{-}(r,\theta)$ is shown along with (b) the amplitude and (c) the phase of the apodized pupil $P(r,\theta)$. (d) An $l=8$ focal plane vortex phase element produces (e) the field at PP2 with $\sim99.98\%$ of the light located outside of the LS.}
\end{figure} 

Large telescopes typically have a central obscuration owing to the secondary mirror and discontinuities formed by spider supports and spaces between mirror segments. The discontinuities cause on-axis starlight to leak into the LS of a VC, thereby limiting the starlight suppression (see e.g. Ref. [10]). To demonstrate how Zernike apodizing functions may be used, we consider a simplified pupil geometry of the future European Extremely Large Telescope (E-ELT) (see Fig. \ref{fig:3}(a)). The pupil of the E-ELT may be characterized by an outer octagon-like mirror and a centrally obscuring hexagonal secondary mirror whose width is $\sim25\%$ of the outer aperture dimension. The primary mirror is constructed of approximately 800 hexagonally packed mirror segments, which we ignore for this discussion. There are six radial spiders with width $\sim3\%$ of the outer aperture dimension. By inspection, the $Z_3^3$ Zernike polynomial apodizer has zero transmission along six radial lines. Naturally, we wish to align these with the spiders of the E-ELT pupil (see Fig. \ref{fig:3}(a)-(c)). The pupil obscurations fit within the dark regions of the apodizer and, therefore, the coronagraph acts similar to the example in Fig. \ref{fig:2}(e). Approximately $99.98\%$ of the on-axis starlight is relocated outside of a circular LS with radius $a_L=0.97a$ if $l\ge8$ and even valued (see Fig. \ref{fig:3}(d)-(e)). For this example, we choose the outer boundary of the apodizer to be a circle slightly smaller that the full E-ELT pupil. That is, the transmission of the apodizer is zero for $r>a$, leading to $\sim5\%$ loss in power. Optimization of the apodizer and Lyot stop boundaries may lead to better performance. This, and the effect of the individual mirror segments, will be considered in future work. 

\section{Performance of Zernike Amplitude Apodized VC's}
In the section, we characterize the performance of Zernike amplitude apodized VC's with particular focus on a $Z_3^3$ apodizer for the E-ELT pupil with an $l=8$ vortex phase element in FP1. 

\subsection{Pupil element transmission}
If a single optical element is used to produce the Zernike apodization over a circular pupil, the relative transmission of the element is given by   
\begin{align}
\tau_{n,m}&=\frac{1}{\pi a^2}\int_0^{2\pi}\int_0^a |P(r,\theta)|^2\,r\,dr\,d\theta, \\
&=\frac{1}{\pi}\int_0^{2\pi}\int_0^1 \left[Z_n^m(\rho,\theta)\right]^2\,\rho\,d\rho\,d\theta,\\
&=\left\{ \begin{matrix}
   \left(n+1\right)^{-1} & m=0  \\
   \left(2n+2\right)^{-1} & m\ne0  \\
\end{matrix} \right. ,
\end{align}
where $\rho=r/a$. The numerical values of $\tau_{n,m}$ for the Zernike amplitude apodized VC's shown in Figs. \ref{fig:2} and \ref{fig:3} are reported in Table 1. Also shown is the relative throughput of the LS, $T^{(0)}_{n,m}/\mathcal{P}$, in each case (see Eq. \ref{eq:throughputdef}). In essence, Zernike amplitude apodizing elements have more loss at higher values of $n$ and $m$. Additional loss is expected owing to the downsized LS; that is, the LS transmission also decreases as $n$ and $m$ increase. 

\begin{table}[t!]
\caption{Performance of Zernike amplitude apodized VC with an unobscured circular pupil. $\tau_{n,m}$ is the apodizer transmission, $T^{(0)}_{n,m}/\mathcal{P}$ is the relative LS throughput for $l=0$ ($a_L=0.97a$), and IWA is the inner working angle.} 
\label{tab:tab1}
\begin{center}       
\begin{tabular}{|c|c|c|c|c|c|} 
\hline
\rule[-1ex]{0pt}{3.5ex}  $n$ & $m$ & $l$ & $\tau_{n,m}$ & $T^{(0)}_{n,m}/\mathcal{P}$ & IWA $\left(\lambda/D\right)$  \\
\hline
\rule[-1ex]{0pt}{3.5ex}  0 & 0 & 2 & 1.00 & 0.94 & 0.9   \\
\hline
\rule[-1ex]{0pt}{3.5ex}  0 & 0 & 4 & 1.00 & 0.94 & 1.6   \\
\hline
\rule[-1ex]{0pt}{3.5ex}  1 & 1 & 4 & 0.25 & 0.89 & 1.3 - 2.1   \\
\hline
\rule[-1ex]{0pt}{3.5ex}  2 & 0 & 4 & 0.33 & 0.84 & 1.1 \\
\hline
\rule[-1ex]{0pt}{3.5ex}  2 & 2 & 6 & 0.17 & 0.83 & 2.2 - 3.2  \\
\hline
\rule[-1ex]{0pt}{3.5ex}  3 & 1 & 6 & 0.13 & 0.81 & 1.7 - 2.3  \\
\hline
\rule[-1ex]{0pt}{3.5ex}  3 & 3 & 8 & 0.13 & 0.78 & 3.2 - 4.0 \\
\hline
\end{tabular}
\end{center}
\end{table} 

\begin{figure}[b!]
\begin{center}
\begin{tabular}{c}
\includegraphics{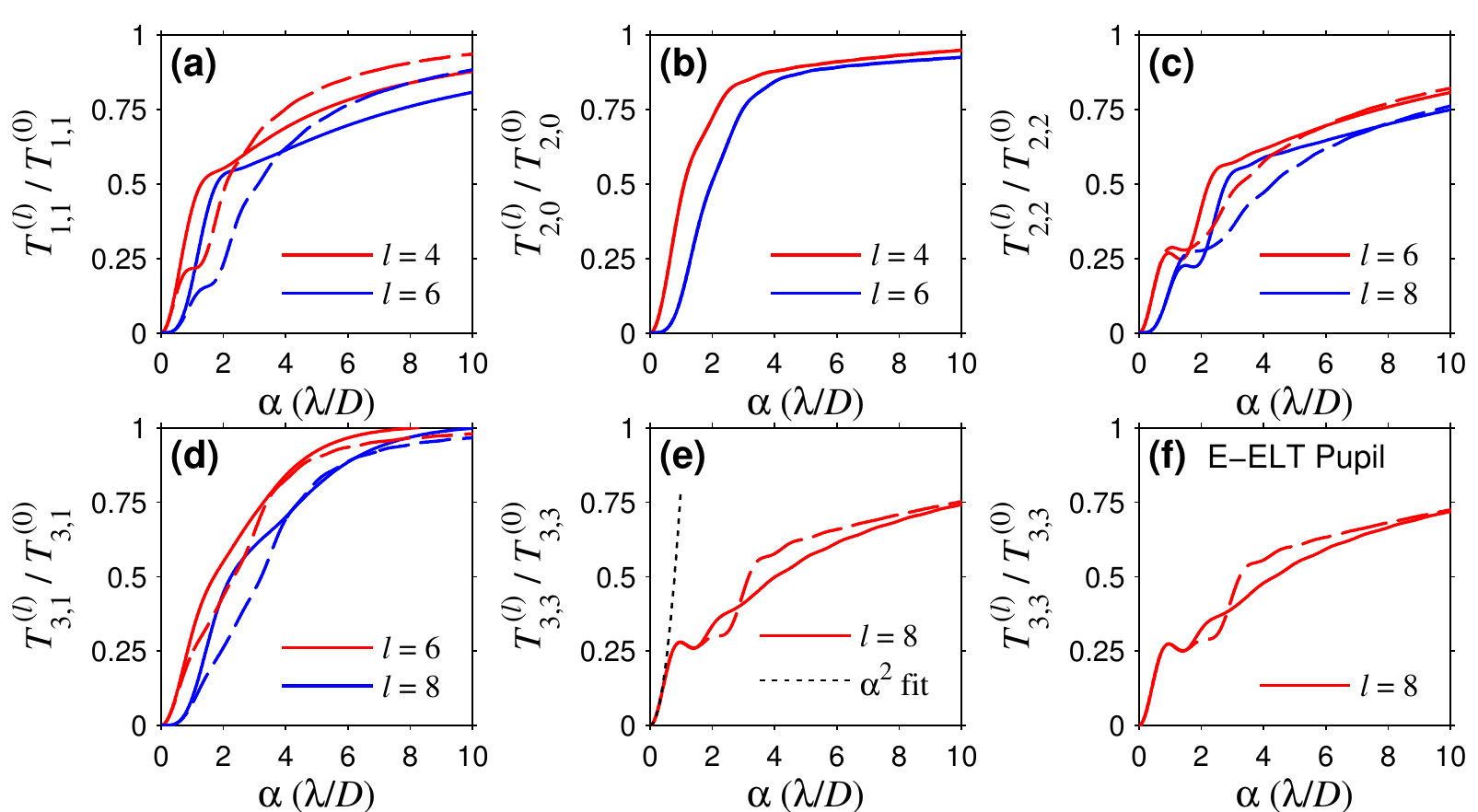}
\end{tabular}
\end{center}
\caption{ \label{fig:4} 
Off-axis throughput for Zernike amplitude apodized VC with an (a)-(e) unobscured circular pupil and (f) the simplified E-ELT pupil. The apodizers considered are (a) $Z_1^1$, (b) $Z_2^0$, (c) $Z_2^2$, (d) $Z_3^1$, and (e) $Z_3^3$. Throughput is plotted with angular displacement $\alpha$ along the vertical (solid) and horizontal (dashed) directions ($45^\circ$ direction is shown as dashed lines in (c)). The throughput for (f) the $Z_3^3$ apodizer applied to the E-ELT pupil is similar to (e). It is also shown in (e) that the throughput in the $Z_3^3$ case increases as $\alpha^2$ for small angles. }
\end{figure} 

\subsection{Off-axis throughput}
The throughput for a point source displaced from the optical axis by angle $\alpha$ is plotted in Fig. \ref{fig:4} for all of the apodizers shown in Figs. \ref{fig:2} and \ref{fig:3}. The throughput curves are calculated by Eq. \ref{eq:throughputdef} and are normalized by the throughput in the $l=0$ case. The exit pupil fields are obtained using fast Fourier transforms on a $4096\times4096$ computational grid with 144 samples over the outer pupil dimension and 28 samples per $\lambda \, F\#$ in the image plane ($F\#=f/D$, where $D=2a$). The off-axis throughput is reduced as compared to the conventional VC without an apodizer (see Fig. \ref{fig:1}(d)). However, we find by inspection that for small angular displacements (i.e. $\alpha \lesssim \lambda/D$), the off-axis throughput increases as $\alpha^x$, where $x=|l|-n-|m|$. For the $Z_3^3$ apodized E-ELT pupil, the off-axis throughput of the $l=8$ VC increases as $\alpha^2$ at small angular displacements, which is comparable to the unobscured $l=2$ system without apodization. Thus, the throughput performance is very good at small angles, especially for a heavily obscured system with thick spiders.

\subsubsection{Inner working angle}
The ``inner working angle" (IWA) is a common measure of the off-axis throughput performance for coronagraphs, typically defined as the angle at which the throughput of a point source is half of its maximum \cite{Guyon2006}. The maximum throughput is achieved when $\alpha \gg \lambda/D$ and may be approximated by the throughput in the $l=0$ case. The IWA's of the Zernike amplitude apodized VC with an unobscured pupil are reported in Table 1. We argue that the IWA is not an optimal performance metric for this system because the off-axis throughput curve for the ELT design has a distinct ``shoulder" that causes the system to have large IWA although the throughput at small angular displacement is relatively high. For example, the throughput reaches $25\%$ of the maximum value at $\alpha=0.7\lambda/D$ for the E-ELT design (compared with $\alpha\approx0.5\lambda/D$ for $l=2$ and $\alpha\approx\lambda/D$ for $l=4$ in the case of an unobscured pupil without apodization).

\begin{figure}[t!]
\begin{center}
\begin{tabular}{c}
\includegraphics{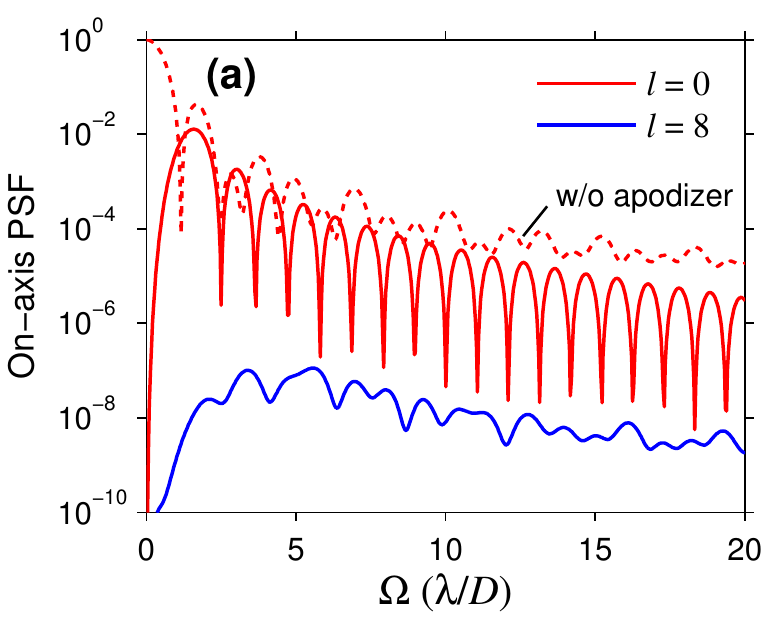}\includegraphics{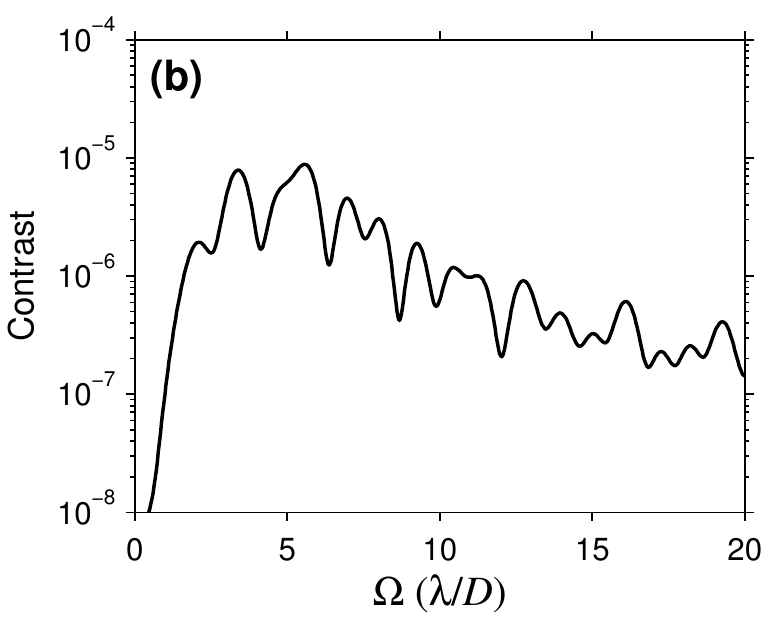}
\end{tabular}
\end{center}
\caption{ \label{fig:5} 
(a) Azimuthally averaged radial profile of the PSF for the E-ELT pupil without (dashed line) and with (solid lines) the $Z_3^3$ pupil apodizer. The post-coronagraph on-axis PSF for an $l=8$ focal plane vortex phase element and $Z_3^3$ apodizer (blue line) estimates the residual starlight in the image plane. $\Omega$ denotes the angular coordinate in the image plane with respect to the optical axis. (b) Azimuthally averaged contrast curve for the apodized coronagraphic system, calculated by normalizing the post-coronagraph on-axis PSF by the peak irradiance of the $l=0$ apodized PSF (or equivalently, the PSF with $\alpha\gg\lambda/D$).}
\end{figure} 

\subsubsection{Post-coronagraph point spread functions}
The residual irradiance owing to the on-axis star dictates the contrast performance of the coronagraph. The azimuthally averaged post-coronagraph on-axis point spread function (PSF) for the E-ELT system is shown in Fig. \ref{fig:5}(a). The PSF in the $l=0$ case with and without the apodizer are shown for comparison. The contrast is defined as the ratio between the residual starlight and the peak planet irradiance in the image plane (see Fig. \ref{fig:5}(b)). We find that the average contrast within $\Omega\le5\lambda/D$ of the star is $2.5\times10^{-6}$, where $\Omega$ is the angular coordinate in the image plane with respect to the $z$ axis. 

\begin{figure}[t!]
\begin{center}
\begin{tabular}{c}
\includegraphics{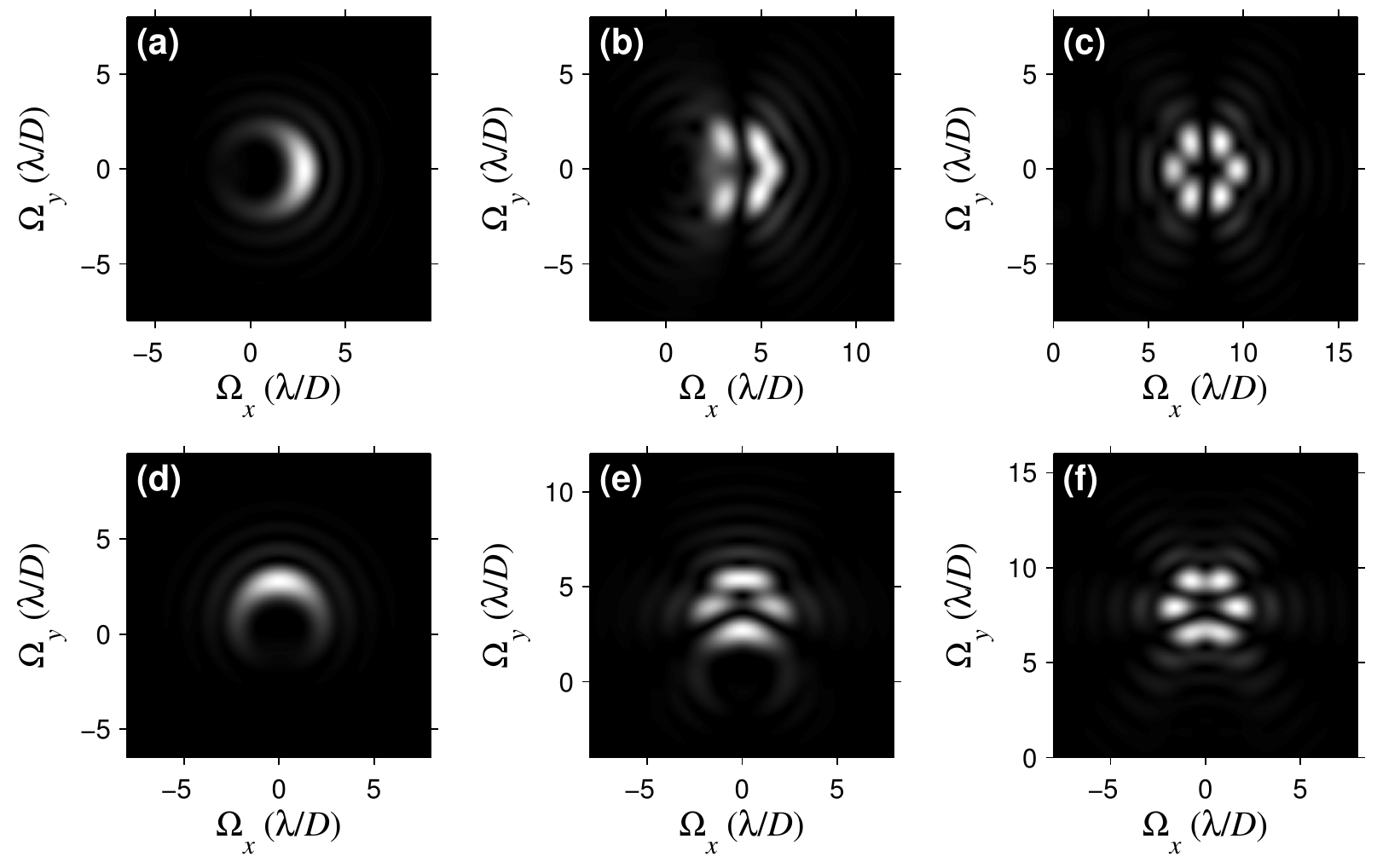}
\end{tabular}
\end{center}
\caption{ \label{fig:6} 
Post-coronagraph off-axis PSF's for angular displacements (a) $\alpha=1.5\,\lambda/D$, (b) $\alpha=4\,\lambda/D$, and (c) $\alpha=8\,\lambda/D$ in the $\Omega_x$ direction. (d)-(f) Same as (a)-(c), but with angular displacements in the $\Omega_y$ direction. $\left(\Omega_x,\Omega_y\right)$ are the angular coordinates with respect to the optical axis. For reference, the peak irradiance values, relative to the maximum of the $l=0$ apodized PSF, are (a) 0.23, (b) 0.23, (c) 0.38, (d) 0.23, (e) 0.31, and (f) 0.37.}
\end{figure}

Like other Lyot-style coronagraphs, the off-axis PSF is spatially-variant. The PSF for a few example angular displacements are shown in Fig. \ref{fig:6}. At small angular displacements, the PSF varies significantly. However, the PSF approaches Eq. \ref{eq:PSF} at $\alpha\gtrsim$ IWA. We note that the PSF's are wider than the VC without an apodizer and, therefore, the angular resolution may be reduced. The effective resolution depends strongly on the post-processing techniques used, a discussion of which is beyond the scope of this work.  

\section{LINEAR COMBINATIONS OF ZERNIKE AMPLITUDE FUNCTIONS}
In this section, we briefly describe how the apodizers presented above may be generalized to linear combinations of Zernike polynomials, while maintaining ideal contrast for unobscured pupils. For a given value of $l$, a linear combination of Zernike amplitude apodization functions of the form 
\begin{equation}
P\left(r,\theta\right) = \sum_{n,m}c_{n,m}Z_n^m\left(r/a,\theta\right),\;\;\;\;\;r\le a,
\label{eq:lincomb}
\end{equation}
yield an nodal area in the on-axis starlight at PP2 provided $l$ is even and $|l|>N+M+1$, where $N$ and $M$ are the maximum values of $n$ and $|m|$, respectively. The coefficients $c_{n,m}$ may be complex constants and each Zernike polynomial may be rotated an arbitrary amount owing to the symmetry of the vortex phase element and nodal area. Although most current implementations of the VC have vortex charge $l=2$ or $l=4$, increasing the value of $l$ allows for many possible apodization functions to be devised to improve performance with complicated pupil obstructions.

Low values of $l$ are typically used in practice because fabricating elements with higher $l$ values is technically challenging. We note, however, that high contrast may also be achieved even for functions that have Zernike coefficients $c_{n',m'}$, where $n'+|m'|+1>|l|$ provided the magnitudes of the lower order Zernike coefficients are relatively large. For an unobscured pupil, the fraction of total starlight power that leaks into the circular nodal area is $\eta=P_{leak}/\mathcal{P}$, where
\begin{equation}
P_{leak}=\sum_{n',m'}\left|c_{n',m'}\right|^2\tau_{n',m'}.
\end{equation}
The sum is taken over all indices $\{n',m'\}$ where $n'+|m'|+1>|l|$. 

In general, the apodizers described in the previous sections may be further optimized for complicated apertures by introducing additional Zernike polynomials with $n+m+1<|l|$. Additionally, Zernike polynomials with $n+m+1>|l|$ may have a negligible effect to the contrast performance if $\left|c_{n,m}\right|^2\ll1$. Optimized apodizers are beyond the scope of this work and will be the topic of a future correspondence. 

\section{CONCLUSION}
We have presented a set of apodizers for the VC that maintain ideal contrast when paired with unobscured circular apertures. The pupil field amplitude is described by Zernike polynomials or linear combinations thereof. These apodizers have features that are particularly useful for high-contrast imaging on ground-based telescopes with thick radial support structures that would otherwise severely limit the starlight suppression capability. Potential drawbacks of this approach include a loss in angular resolution owing to the widening of the PSF as well as reduced transmission. To mitigate the loss in the system and improve off-axis performance, we propose that this approach is used in combination with phase induced apodization techniques \cite{Guyon2003,Pueyo2013,Pueyo2014,Fogarty2014}. Our design for the E-ELT requires an $l=8$ vortex phase element, which is not typically used in practice due to fabrication challenges. However, the on-axis zero in the pre-coronagraph PSF is expected to reduce the sensitivity to central defects in the vortex phase element as compared to the conventional VC without a pupil apodizer or obscurations. Future work will investigate the implementation of lossless Zernike amplitude apodization and the reduced sensitivity to defects in the vortex phase element. In addition, we will implement relaxation techniques to smooth amplitude discontinuities in the pupil function when matched with heavily obscured apertures and explore potential contrast improvements via interferometry techniques \cite{Riaud2014}. We expect this work to improve on methods used to find optimized apodizers for complicated pupils by providing a new analytical basis for analysis \cite{Mawet2013b,Mawet2013c,Carlotti2011,Carlotti2013,Carlotti2014}. 

\appendix    
\section{Zernike Polynomials} \label{sec:Zernikes}
The Zernike polynomials may be written as 
\begin{equation}
Z_n^m(\rho,\theta) = R_n^m(\rho)\cos(m\theta),\;\;\;\;\;\rho\le1,
\end{equation}
\begin{equation}
Z_n^{-m}(\rho,\theta) = R_n^m(\rho)\sin(m\theta),\;\;\;\;\;\rho\le1,
\end{equation}
where $\rho=r/a$ and $R_n^m(\rho)$ is the radial Zernike polynomial given by 
\begin{equation}
R_n^m(\rho) = \sum_{k=0}^{\frac{n-m}{2}}\frac{\left(-1\right)^k\left(n-k\right)!}{k!\left(\frac{n+m}{2}-k\right)!\left(\frac{n-m}{2}-k\right)!}\rho^{n-2k},\;\;\;\;\;\rho\le1,
\end{equation}
where $n-m$ is even. The indices $n$ and $m$ are integers respectively known as the ``degree" and ``azimuthal order." The first few radial polynomials are: $R_0^0=1$, $R_1^1=\rho$, $R_2^0=2\rho^2-1$, $R_2^2=\rho^2$, $R_3^1=3\rho^3-2\rho$, $R_3^3=\rho^3$.

\section{Expressions for the fields at the exit pupil} \label{sec:EPfields}
The exit pupil field patterns for the first few Zernike amplitude apodizers (i.e. solutions to Eqn. \ref{eq:Wevens}) are shown in Fig. \ref{fig:2} and are written below:
\begin{equation}
g^{(4)}_{1,1}\left(r'',\psi \right)=\left[2\left(\frac{a}{r''}\right)^5-\frac{3}{2}\left(\frac{a}{r''}\right)^3\right]{e^{i5\psi}}+\frac{1}{2}\left(\frac{a}{r''}\right)^3{e^{i3\psi}},
\end{equation}
\begin{equation}
g^{(4)}_{2,0}\left(r'',\psi \right)=\left(\frac{a}{r''}\right)^4{e^{i4\psi}},
\end{equation}
\begin{equation}
g^{(6)}_{2,2}\left(r'',\psi \right)= \left[\frac{21}{2}\left(\frac{a}{r''}\right)^8 -15\left(\frac{a}{r''}\right)^6 + 5\left(\frac{a}{r''}\right)^4 \right]e^{i8\psi}+\frac{1}{2}\left(\frac{a}{r''}\right)^4 e^{i4\psi},
\end{equation}
\begin{equation}
g^{(6)}_{3,1}\left(r'',\psi \right)=\left[3\left(\frac{a}{r''}\right)^7-\frac{5}{2}\left(\frac{a}{r''}\right)^5\right]{e^{i7\psi}}+\frac{1}{2}\left(\frac{a}{r''}\right)^5{e^{i5\psi}},
\end{equation}
\begin{equation}
g^{(8)}_{3,3}\left(r'',\psi \right)=\left[60\left(\frac{a}{r''}\right)^{11}-126\left(\frac{a}{r''}\right)^9+84\left(\frac{a}{r''}\right)^7-\frac{35}{2}\left(\frac{a}{r''}\right)^5\right]{e^{i11\psi}}+\frac{1}{2}\left(\frac{a}{r''}\right)^5{e^{i5\psi}}.
\end{equation}

\acknowledgments     
This work was supported by the National Science Foundation under Grant No. ECCS-1309517.

\end{document}